\documentclass[twocolumn]{aastex}
\usepackage[english]{babel}
\usepackage{amsmath}
\usepackage{graphicx}
\usepackage{wasysym}
\usepackage{soul}
\usepackage{parskip}
\usepackage{wrapfig}
\usepackage{url}
\usepackage{float}
\usepackage{hyperref}
\usepackage{xcolor}

\newcommand{\tess}{\emph{TESS}}

\colorlet{Mycolor1}{green!72!red!228!}

\shorttitle{Gravity-Darkening Analysis of KELT-9 b}
\shortauthors{Ahlers et al.}
\graphicspath{{./}{figures/}}

\begin{document}

\title{KELT-9 b's Asymmetric \tess\ Transit Caused by Rapid Stellar Rotation and Spin-Orbit Misalignment}

\correspondingauthor{John P. Ahlers}
\email{johnathon.ahlers@nasa.gov}

\author[0000-0003-2086-7712]{John P. Ahlers}
\altaffiliation{NASA Postdoctoral Program Fellow}
\affiliation{Exoplanets and Stellar Astrophysics Laboratory, Code 667, NASA Goddard Space Flight Center (GSFC), Greenbelt, MD 20771, USA}
\affiliation{GSFC Sellers Exoplanet Environments Collaboration}

\author{Marshall C. Johnson}
\affiliation{Las Cumbres Observatory, 6740 Cortona Drive, Suite 102, Goleta, CA 93117, USA}


\author{Keivan G. Stassun}
\affiliation{Vanderbilt University, Department of Physics \& Astronomy, 6301 Stevenson Center Lane, Nashville, TN 37235, USA}
\affiliation{Fisk University, Department of Physics, 1000 17th Avenue N., Nashville, TN 37208, USA}

\author{Knicole D. Col\'on}
\affiliation{Exoplanets and Stellar Astrophysics Laboratory, Code 667, NASA Goddard Space Flight Center, Greenbelt, MD 20771, USA}
\affiliation{GSFC Sellers Exoplanet Environments Collaboration}

\author{Jason W. Barnes}
\affiliation{Department of Physics, University of Idaho, Moscow ID 83844-0903 USA}

\author[0000-0002-5951-8328]{Daniel J. Stevens}
\altaffiliation{Eberly Fellow}
\affiliation{Department of Astronomy \& Astrophysics, The Pennsylvania State University, 525 Davey Lab, University Park, PA 16802, USA}
\affiliation{Center for Exoplanets and Habitable Worlds, The Pennsylvania State University, 525 Davey Lab, University Park, PA 16802, USA}

\author{Thomas Beatty}
\affiliation{Department of Astronomy and Steward Observatory, University of Arizona, Tucson, AZ 85721, USA}

\author{B. Scott Gaudi}
\affiliation{Department of Astronomy, The Ohio State University, 140 West 18th Avenue, Columbus, OH 43210, USA}

\author[0000-0001-6588-9574]{Karen A.\ Collins}
\affiliation{Center for Astrophysics \textbar \ Harvard \& Smithsonian, 60 Garden Street, Cambridge, MA 02138, USA}

\author{Joseph Rodriguez}
\affiliation{Center for Astrophysics \textbar \ Harvard \& Smithsonian, 60 Garden Street, Cambridge, MA 02138, USA}

\author{George Ricker}
\affiliation{Department of Physics and Kavli Institute for Astrophysics and Space Research, Massachusetts Institute of Technology, Cambridge, MA 02139, USA}

\author{Roland Vanderspek}
\affiliation{Department of Physics and Kavli Institute for Astrophysics and Space Research, Massachusetts Institute of Technology, Cambridge, MA 02139, USA}

\author{David Latham}
\affiliation{Center for Astrophysics \textbar \ Harvard \& Smithsonian, 60 Garden Street, Cambridge, MA 02138, USA}

\author{Sara Seager}
\affiliation{Department of Physics and Kavli Institute for Astrophysics and Space Research, Massachusetts Institute of Technology, Cambridge, MA 02139, USA}
\affiliation{Department of Earth, Atmospheric and Planetary Sciences, MIT, Cambridge, MA 02139, USA}
\affiliation{Department of Aeronautics and Astronautics, MIT, Cambridge, MA 02139, USA}

\author{Joshua Winn}
\affiliation{Department of Astrophysical Sciences, Princeton University, Princeton, NJ 08544, USA}

\author{Jon M. Jenkins}
\affiliation{NASA Ames Research Center, Moffett Field, CA 94035, USA}

\author{Douglas A. Caldwell}
\affiliation{NASA Ames Research Center, Moffett Field, CA 94035, USA}

\author{Robert F. Goeke}
\affiliation{Department of Physics and Kavli Institute for Astrophysics and Space Research, Massachusetts Institute of Technology, Cambridge, MA 02139, USA}



\author{Hugh P. Osborn}
\affiliation{Department of Physics and Kavli Institute for Astrophysics and Space Research, Massachusetts Institute of Technology, Cambridge, MA 02139, USA}
\affiliation{NCCR/PlanetS, Centre for Space \& Habitability, University of Bern, Bern, Switzerland}

\author{Martin Paegert}
\affiliation{Center for Astrophysics \textbar \ Harvard \& Smithsonian, 60 Garden Street, Cambridge, MA 02138, USA}

\author{Pam Rowden}
\affiliation{School of Physical Sciences, The Open University, Milton Keynes MK7 6AA, UK}

\author{Peter Tenenbaum}
\affiliation{NASA Ames Research Center, Moffett Field, CA 94035, USA}
\affiliation{SETI Institute, 189 Bernardo Avenue, Suite 200, Mountain View, CA 94043, USA}



\begin{abstract}
KELT-9 b is an ultra hot Jupiter transiting a rapidly rotating, oblate early-A-type star in a polar orbit. We model the effect of rapid stellar rotation on KELT-9 b's transit light curve using photometry from the Transiting Exoplanet Survey Satellite (\tess) to constrain the planet's true spin-orbit angle and to explore how KELT-9 b may be influenced by stellar gravity darkening. We constrain the host star's equatorial radius to be $1.089\pm0.017$ times as large as its polar radius and its local surface brightness to vary by $\sim38$\%  between its hot poles and cooler equator. We model the stellar oblateness and surface brightness gradient and find that it causes the transit light curve to lack the usual symmetry around the time of minimum light. We take advantage of the light curve asymmetry to constrain KELT-9 b's true spin orbit angle (${87^\circ}^{+10^\circ}_{-11^\circ}$), agreeing with \citet{gaudi2017giant} that KELT-9 b is in a nearly polar orbit. We also apply a gravity darkening correction to the spectral energy distribution model from \citet{gaudi2017giant} and find that accounting for rapid rotation gives a better fit to available spectroscopy and yields a more reliable estimate for the star's polar effective temperature.  
\end{abstract}


\section{Introduction}\label{sec:intro}

KELT-9 b (TIC 16740101) is one of the hottest confirmed planets to date. The  $2.88\pm0.35M_\mathrm{Jup}$ planet orbits a B9.5-A0 star in a 1.48-day orbit, with an estimated dayside equilibrium temperature of $\sim4600$ K \citep{gaudi2017giant,Cauley2019,kitzmann2018peculiar,hoeijmakers2019spectral}. The host star HD 195689 (hereafter called KELT-9) is more than twice the radius of the Sun and has an effective temperature of roughly 10,000 K, making it $\sim50$ times more luminous. At any given time, KELT-9 b receives $\sim$44,000 times as much incident flux as the Earth. \\

KELT-9 b's equilibrium temperature and insolation are more complicated than previously assumed because of its host star's rapid rotation \citep{gaudi2017giant}. KELT-9's high internal angular momentum ($v\sin(i)=111.4\pm1.3$ km/s)  flattens it into an oblate spheroid, making the equatorial radius of the star larger than the polar radius. Additionally, the star's abundant centrifugal force near its equator distorts its hydrostatic equilibrium, causing its effective temperature to vary by nearly a thousand Kelvin over the surface of the star. These two effects of stellar oblateness and varying effective temperature -- together commonly referred to as gravity darkening \citep{barnes2009transit} -- change the total irradiance on KELT-9 b \citep{ahlers2016gravity}. The star's oblateness changes the overall shape and size of the projected disk in the sky that KELT-9 b sees depending on its location in the system, and the star's decreased temperature decreases output stellar radiation near its equator. \\

The effect of gravity-darkening on KELT-9 b is compounded by the planet's orbit. Using the 1.5 m Tillinghast reflector and TRES spectrograph, \citet{gaudi2017giant} first measured KELT-9 b's projected alignment angle to be $-84.8^\circ\pm1.4^\circ$ via Doppler tomography, meaning KELT-9 b resides in a polar orbital configuration. Therefore, KELT-9 b varies in exposure to the host star's hotter poles and cooler equator, which has been shown to significantly impact a planet's total irradiation in similar systems \citep{ahlers2016gravity,ahlers2020gravity}. \\

KELT-9 b is an especially interesting target for hot Jupiter research. In its polar orbit, KELT-9 b follows the trend that gas giants around high-mass stars are frequently spin-orbit misaligned \citep[e.g.,][]{winn2010hot,2012ApJ...757...18A,winn2015occurrence,zhou2019two}. Its high dayside temperature provides excellent opportunities for phase curve and secondary eclipse analyses \citep{hooton2018ground,wong2019exploring,mansfield2020evidence}. The high signal-to-noise ratio of its transit makes KELT-9 b a top target for transmission spectroscopy \citep{hoeijmakers2018atomic,Cauley2019,hoeijmakers2019spectral}. In this work, we model KELT-9 b's \tess~light curve including rapid stellar rotation to measure the hot Jupiter's transit parameters, and we take advantage of the transit asymmetry caused by gravity darkening to robustly constrain the planet's orbital geometry including its true spin-orbit orientation.

This work marks one of only a handful to account for gravity-darkening in an exoplanet analysis \citep{2011ApJS..197...10B,zhou2013highly,ahlers2014,ahlers2015spin,masuda2015spin,barnes2015probable,ahlers2019dealing,ahlers2020gravity}. In \S \ref{sec:methods} we describe the gravity-darkening technique we use to measure KELT-9 b's alignment angle. In \S \ref{sec:results} we show the results of our \tess~photometry analysis. In \S \ref{sec:discussion} we discuss possible causes of KELT-9 b's polar orbit and address the effects stellar gravity-darkening can have on the hot Jupiter's atmospheric processes and equilibrium temperature.

\section{Methods}\label{sec:methods}

\subsection{\tess~Photometry} \label{sec:photometry}

\begin{figure*}
\includegraphics[width=\textwidth]{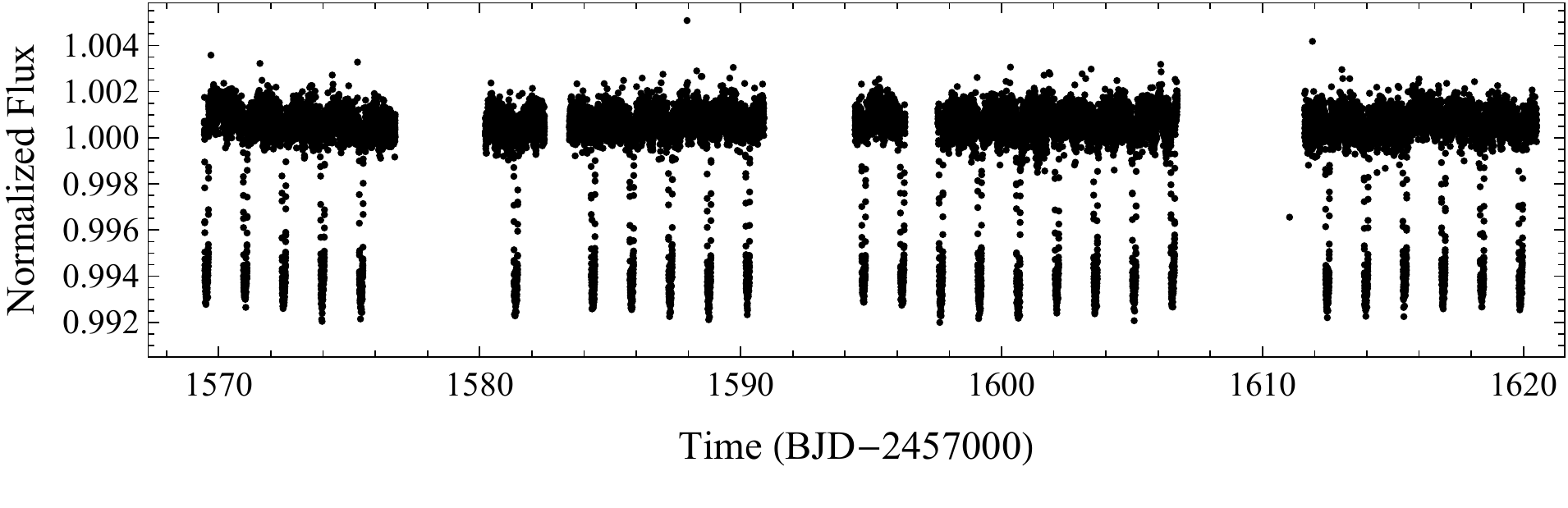}
\caption{\tess~observed 26 complete transits during Sectors 14 and 15, and one partial transit at $\sim1611$ days. We do not include the partial transit in our analysis. The dataset includes clear secondary eclipses and phase curve signals; we focus on primary transit events in this work. The full detrended light curve is shown above.}
\label{fig:tsplot}
\end{figure*}

\tess~observed 27 transits of KELT-9 b at 2-minute cadence in sectors 14 and 15 from July 18, 2019 to September 11, 2019 during the northern observing campaign as part of two Guest Investigator Programs: 22197 and 22053.  The light curves produced by the TESS Science Processing Operations Center (SPOC) were downloaded from the Mikulski Archive for Space Telescopes. 

The available \tess~photometry of KELT-9 b is broken up into four 13.5-day segments due to \tess's orbit. The fourth time series segment began with a partial transit, which we removed from our dataset, leaving 26 transits for our analysis. We apply a 36-hour median box filter to the photometry to correct for long-term systematics in each segment. We show the full normalized light curve in Figure \ref{fig:tsplot}. We phase-fold the light curve on KELT-9 b's orbital period and re-bin at 120 seconds to reduce computation time, following previous gravity-darkening transit analyses \citep[e.g.,][]{ahlers2020gravity}.

\subsection{Spin-Orbit Angle} \label{sec:gdark}
The primary goal of this analysis is to measure KELT-9 b's spin-orbit angle from \tess~photometry. We take advantage of the host star KELT-9's rapid rotation and apply the gravity-darkening transit model \citep{von1924radiative,barnes2009transit}, which measures both the stellar inclination and the projected orbital inclination for a transiting planet. See Figure \ref{fig:bestfit} for definitions of these angles.  

KELT-9 is a rapid rotator at its surface with $v\sin(i)=111.4\pm1.3~\mathrm{km/s}$ \citep{gaudi2017giant}, which induces two effects. First, the star flattens into an oblate shape due to the high centrifugal force near its equator. Second, the star's surface gravity decreases near the equator, resulting in an effective temperature gradient that varies nearly 1000 K between the hot poles and cooler equator. We include both the oblateness and the temperature gradient in our gravity-darkened model. We show in Figure \ref{fig:SED} the difference in sky-projected spectral energy distributions between gravity-darkened and non-gravity-darkened models of KELT-9.

When KELT-9 b transits its host star, it blocks a certain amount of light depending on whether it is transiting near the bright poles or dim equator. In the case of a misaligned orbit, the planet blocks varying intensities of light throughout its transit, resulting in an asymmetric transit. KELT-9 b's transit ingress is deeper than its egress, meaning the planet begins its transit near KELT-9's hot pole and moves toward its cooler equator.

Taking the host star's oblateness and luminosity gradient into account, the planet's spin-orbit angle can be measured directly from its transit light curve using the gravity-darkening model. Following \citet{barnes2009transit}, we measure both the star's inclination angle and the angle between the sky projections of the planet's orbital axis and the star's rotation axis, which yields the three-dimensional spin-orbit angle $\varphi$ via

\begin{equation}
\cos(\varphi)=\sin(i_\star)\cos(i)+\cos(i_\star)\sin(i)\cos(\lambda),
\label{eq:soangle}
\end{equation}

where $\varphi$ is the true spin-orbit angle, $\lambda$ is the projected orbital alignment, $i_\star$ is the stellar inclination, and $i$ is the orbital inclination. 

\subsection{Modelling Gravity-Darkening and Limb-Darkening}
In this work, we fit the gravity-darkening exponent $\beta$ and both quadratic limb-darkening coefficients in our transit light curve analysis. We numerically integrate the star's asymmetric disk and subtract flux blocked by the planet at every time bin. We apply the Levenberg-Marquardt $\chi^2$ minimization technique to find a best-fit to the \tess\ dataset, following previous gravity darkening transit analyses \citep[e.g.,][]{barnes2009transit}. We start with a theoretical value for $\beta$ from \citet{lara2011gravity} and quadratic limb-darkening coefficients from \citet{claret2017limb} as initial guesses in our fits. Previous gravity-darkening works have used fixed gravity-darkening and limb-darkening values in their transit light curve models, primarily due to those values being difficult to fit. For example, \citet{2011ApJS..197...10B} and \citet{masuda2015spin} showed for the Kepler-13A system that incorrect limb-darkening values can significantly skew spin-orbit angle results. Similarly, \citet{ahlers2014} and \citet{zhou2013highly} obtained nearly opposite answers for KOI-368's spin-orbit angle because of their differing values of the gravity-darkening exponent $\beta$. 

The discrepancies between these works can largely be attributed to the host stars KOI-368 and Kepler-13A being previously uncharacterized at the time of the transit light curve analyses. KELT-9 b is different from those previous works because it is a well-characterized star with significant archival photometry and spectroscopy. With tight constraints on the previously-reported parameters listed in Tables \ref{table:star} and \ref{table:bestfit}, we explored KELT-9's asymmetric light curve in greater detail than ever before for any planetary system, yielding direct gravity-darkening and limb-darkening parameters. 

\renewcommand{\arraystretch}{1.2}
\begin{table*}[htbp]
\centering
\begin{tabular}{l l c r}
\hline \hline {\bf Parameter} & {\bf Description} & {\bf Value} & {\bf Source} \\ \hline
$P$ & orbital period (days) & $1.4811235\pm0.0000011$ & \citet{gaudi2017giant}  \\
$T_\mathrm{eff}$ & stellar effective temperature (K)  & $10170\pm450$ &  \citet{gaudi2017giant}  \\
$M_\star$ & stellar mass ($M_\odot$) & $2.52^{+0.25}_{-0.20}$ &  \citet{gaudi2017giant}  \\
$R_\star$ & stellar radius ($R_\odot$) & $2.362^{+0.075}_{-0.063}$ &  \citet{gaudi2017giant}  \\
$v\sin(i)$ & projected rotational velocity (km/s) & $111.4\pm1.3$ &  \citet{gaudi2017giant}  \\
$\beta$ & gravity-darkening exponent & $0.2\pm0.04$ & \citet{claret2016theoretical} \\
$a$ & first limb-darkening term & 0.1588 & \citet{claret2017limb} \\
$b$ &  second limb-darkening term & 0.2544 & \citet{claret2017limb} \\ 

\end{tabular}
\caption{Previously-reported or theoretical values for the KELT-9 system relevant to our transit analysis. We adopt $T_\mathrm{eff}$, $M_\star$, and $v\sin(i)$ and their uncertainties from  \citet{gaudi2017giant} as assumed values. We use the observed $R_\star$ and the theoretical $\beta$, $a$, and $b$ values as initial guesses in our fitting model.}
\label{table:star}
\end{table*}

\section{Results} \label{sec:results}

\begin{figure*}
\begin{tabular}{l r}
\raisebox{0.85cm}{\includegraphics[width=0.47\textwidth]{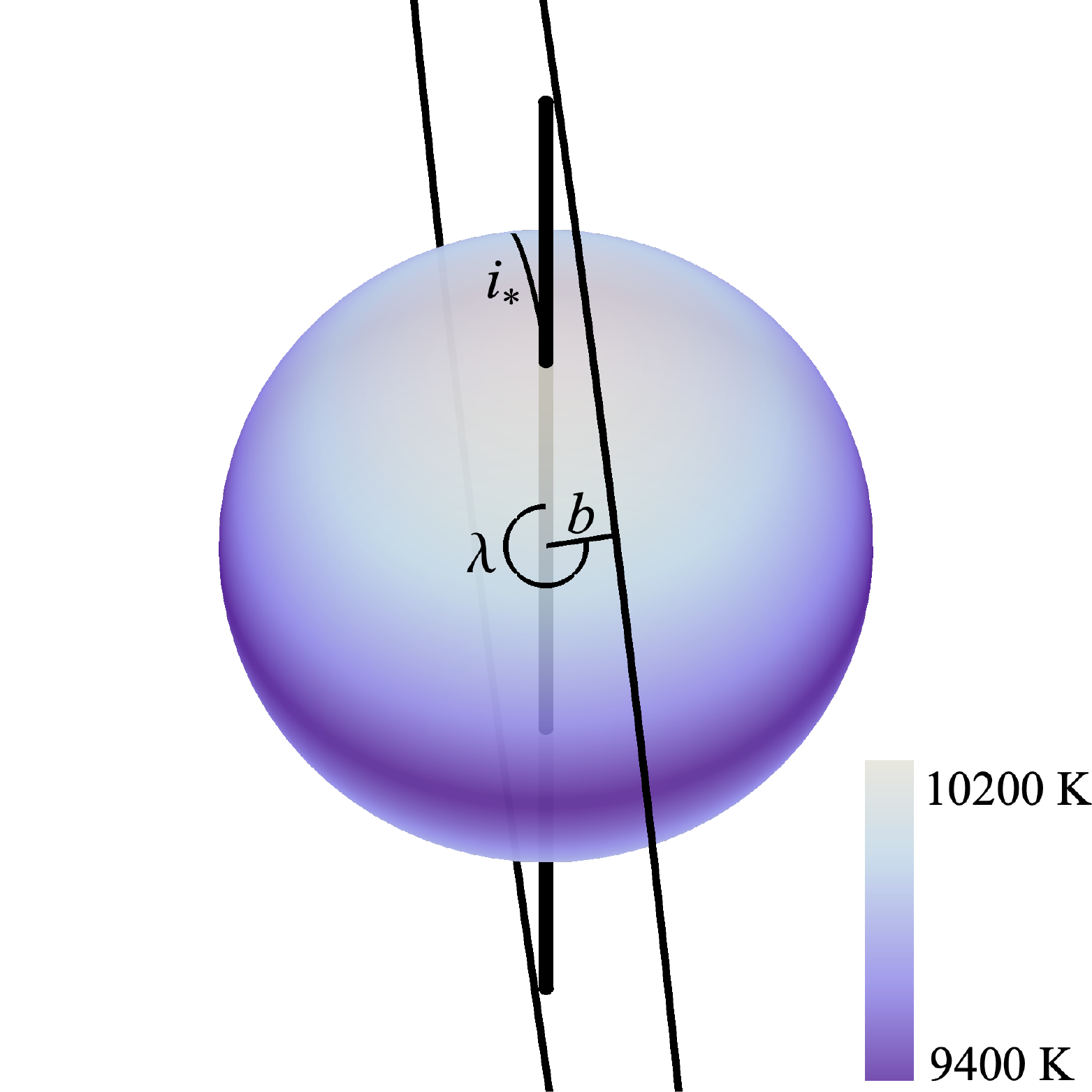}} & \includegraphics[width=0.47\textwidth]{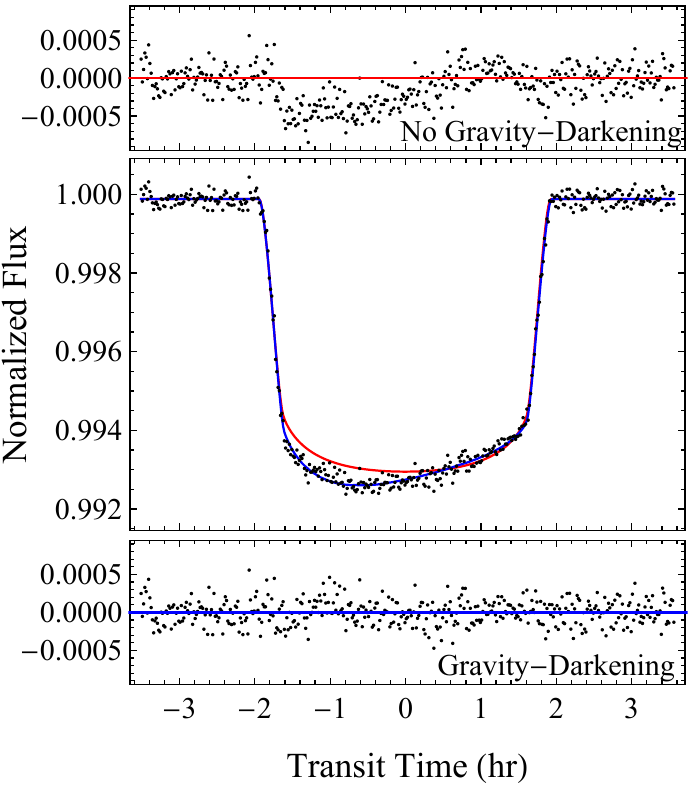} 
\end{tabular}
\caption{(Left) KELT-9 b begins its transit near the star's hot pole and moves toward the star's cooler equator. Our transit analysis directly measures the stellar inclination ($i_\star$), the planet's projected alignment ($\lambda$), and the orbital inclination (i.e., the impact parameter $b$). We find that KELT-9 varies in effective temperature by $\sim800$ K between its hot poles and cooler equator. (Right) KELT-9 b's phase-folded primary transit from \tess. The transit depth steadily decreases throughout the eclipse, indicating that KELT-9 b begins its transit near one of the host star's hotter poles and moves toward the dimmer stellar equator.}
\label{fig:bestfit}
\end{figure*}

\renewcommand{\arraystretch}{1.2}
\begin{table*}[htbp]
\centering
\begin{tabular}{l l c c c}
 \hline \hline {\bf Parameter} & {\bf Description} & {\bf G-Dark} & {\bf No G-Dark} & {\bf Gaudi et al. (2017)} \\ \hline
$\chi^2_\mathrm{red}$ & goodness of fit & 1.30 & 2.8 & 2.56 \\
$R_\star$ & equatorial stellar radius $(R_\odot)$ & $2.39\pm0.03$ & --- & $2.36^{+0.08}_{-0.06}$  \\
$R_\mathrm{p}$ & planet radius $(R_\mathrm{Jup})$ & $1.84\pm0.04$ & --- & $1.89^{+0.07}_{-0.04}$  \\
$R_\mathrm{p}/R\star$ & radii ratio & $0.081\pm0.002$ & $0.079\pm0.003$ & $0.0822\pm0.0004$ \\
$T_0$ & transit epoch (BJD-2457000)  & $1683.4449\pm0.00008$ & $1683.445\pm0.00013$ & --- \\
$i$ & orbital inclination (deg) & $87.2\pm0.4$  & $87.1\pm0.5$ & $86.7\pm0.3$ \\
$i_\star$ & stellar inclination (deg) &  ${52^\circ}^{+8^\circ}_{-7^\circ}$ & --- & --- \\
$\lambda$ & projected alignment (deg) & $-88^\circ\pm 15^\circ$ & --- & $-84.8^\circ\pm1.4^\circ$  \\
$\varphi$ &  spin-orbit angle (deg) & ${87^\circ}^{+10^\circ}_{-11^\circ}$  & --- & --- \\
$P_\star$ & stellar rotation period (hr) & $16^{+5}_{-4}$ & --- & ---  \\\
$\zeta$ & stellar oblateness  & $0.089\pm0.017$ & --- & --- \\
$\beta$ & gravity-darkening exponent & $0.137\pm0.014$ & --- & ---\\ \hline

\hline
\end{tabular}
\caption{Best-fit values from our gravity-darkened model of KELT-9 b's \tess~photometry. We find that the primary transit is dramatically influenced by KELT-9's rapid rotation in \tess's bandpass and that only by accounting for gravity darkening can we achieve a good quality fit. The $\chi^2_\mathrm{red}$ from \citet{gaudi2017giant} is their goodness-of-fit for the SED rather than \tess\ photometry. We define the orbit geometry angles in Figure \ref{fig:bestfit}.}
\label{table:bestfit}
\end{table*}

We apply the gravity darkening model to KELT-9 b's \tess~primary transit to measure the planet's true spin-orbit angle, as well as to determine fundamental parameters about the KELT-9 host star. Figure \ref{fig:bestfit} shows the best-fit to our photometry as well as a transit diagram. Table \ref{table:bestfit} lists our best-fit parameters. 

\subsection{Spin-Orbit Angle}
We measure KELT-9 b's spin orbit angle to be ${87^\circ}^{+10^\circ}_{-11^\circ}$. Using the gravity-darkening approach detailed in \S \ref{sec:gdark}, we measured both the star's axial tilt in/out of the plane of the sky -- i.e. the stellar inclination -- and the star's axial tilt relative to the planet's projected orbital path -- i.e. the projected alignment. Together with the transit impact parameter -- i.e. orbital inclination -- we determine KELT-9 b's true spin-orbit angle via Equation \ref{eq:soangle}.

The gravity-darkening technique cannot distinguish between a retrograde/prograde projected alignment orientation. Fortunately, Doppler tomography uniquely measures projected alignment. \citet{gaudi2017giant} measured $\lambda=-84.8^\circ\pm1.4^\circ$ via Doppler tomography, eliminating the prograde/retrograde degeneracy and yielding a single, robust solution for KELT-9 b's true spin-orbit angle.

\subsection{Stellar Parameters} \label{sec:starmodel}
From \tess's primary transit observations, we constrain the KELT-9 host star's inclination, rotation rate, oblateness, gravity-darkening exponent, and effective temperature gradient. We measure KELT-9's stellar inclination $i_\star={52^\circ}^{+8^\circ}_{-7^\circ}$ directly in our transit best-fit model, defined as $i_\star=0^\circ$ when the star is viewed equator-on, and $i^\star=|90^\circ|$ when viewed pole-on. Combining $i_\star$ with the the star's projected velocity from \citet{gaudi2017giant} ($v\sin(i)=111.4\pm1.3$ km/s), we determine the star's true rotation period at $P_\star=16^{+5}_{-4}$ hours via,
\begin{equation}
P_\star=\frac{2\pi R_\mathrm{eq}\cos(i_\star)}{v\sin(i)}
\end{equation} 

where $R_\mathrm{eq}$ is the star's equatorial radius. 

With KELT-9's rotation rate constrained, we estimate its oblateness using the Darwin-Radau relation \citep{murray1999solar}. In previous works, we estimated the gravity darkening exponent $\beta$ following \citet{claret2016theoretical}. However, with the excellent photometric precision and dramatic transit asymmetry from the \tess\ light curve of KELT-9 b, we instead use the estimated value from \citet{claret2016theoretical} as an initial guess and fit for the gravity darkening exponent directly, obtaining $\beta=0.137\pm0.014$.

We estimate KELT-9's effective temperature across its surface using \citet{gaudi2017giant}, with $T_\mathrm{eff}=10170\pm450$ as the polar effective temperature and modeling the temperature at a given latitude $\theta$ via the von Zeipel theorem,

\begin{equation}
T(\theta)=T_\mathrm{pole}\left(\frac{g(\theta)}{g_\mathrm{pole}}\right)^\beta
\end{equation}

where $g$ is the effective surface gravity. We model stellar surface gravity to second order including a centrifugal force term, following, e.g., \citet{ahlers2016gravity}. 

\subsection{Gravity Darkening SED Correction}
\begin{figure}[htbp]
\includegraphics[width=0.47\textwidth]{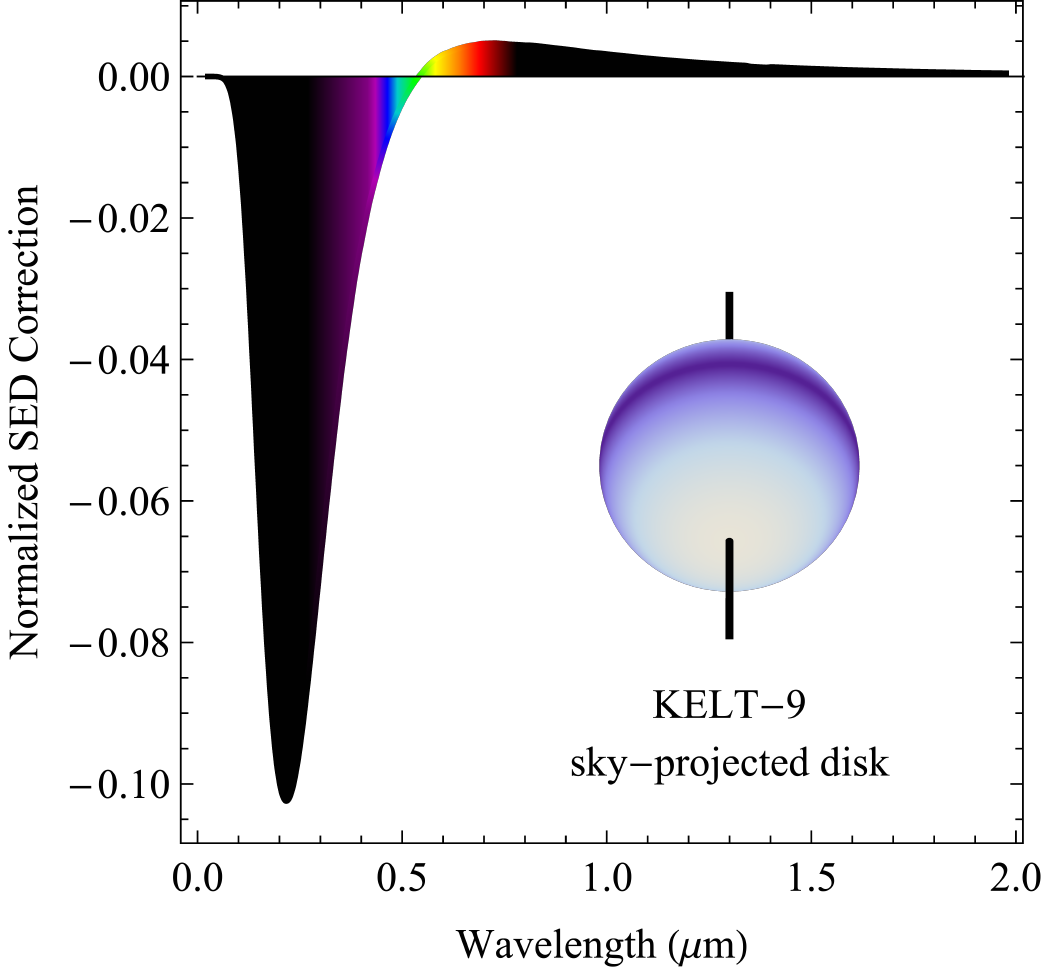} 
\caption{\footnotesize KELT-9's oblateness and gravity-darkening gradient produce an SED that is shifted significantly down in the ultraviolet and slightly up in the visible and infrared compared to a slow rotator of equivalent size and temperature. This plot illustrates the normalized difference between a gravity-darkened and traditional SED as seen in the plane of the sky using our measured and assumed stellar parameters. Gravity darkening decreases the star's output in the violet and near UV by $\sim10$\%, which significantly impacts the irradiance on KELT-9 b.}
\label{fig:SED}
\end{figure}

Using KELT-9's oblateness and temperature gradient (see \S \ref{sec:starmodel}), we model the effect of gravity darkening on KELT-9's spectral energy distribution (SED). We integrate the stellar surface and compare KELT-9's spectrum arising from the sky projected disk with and without rapid rotation. Gravity darkening causes a lower stellar flux in the near UV and blue end of the visible spectrum, and causes a slightly increased flux in the red end of the visible spectrum and IR. Figure \ref{fig:SED} shows the normalized difference that rapid rotation makes in our stellar model. 

We apply the values of Figure \ref{fig:SED} to the \citet{gaudi2017giant} SED model as a gravity darkening correction and improve the SED best-fit result ($\chi^2_\mathrm{red}=1.91$ versus $\chi^2_\mathrm{red}=2.56$ originally). Notably, the corrected model yields $T_\mathrm{eff}=10250\pm250$ K versus $9650\pm550$ K originally, which is a substantially better match with the adopted fiducial model in \citet{gaudi2017giant}. Therefore, the corrected SED model supports our assumptions for stellar mass and temperature (see Table \ref{table:star}). The values we adopt for $M_\star$ and $T_\mathrm{eff}$ agree with both the fiducial model and the corrected SED model in \citet{gaudi2017giant}.

This result demonstrates for the first time that a straightforward gravity darkening correction can improve a standard SED model for a rapidly rotation B/A star and further validates the gravity darkening model that we adopt in this work.

\subsection{Nodal Precession}
In order to estimate the expected nodal precession rate of KELT-9 b, we essentially followed the methodology that \citet{iorio2011classical} introduced for the hot Jupiter WASP-33~b, updated to account for the fact that for KELT-9 we have measured many of the parameters that \citet{iorio2011classical} had to estimate. This assumes that the stellar spin angular momentum is significantly larger than the planetary orbital momentum, and neglects contributions to the precession rate from general relativity and any additional objects in the KELT-9 system. 

The major parameter that we need to estimate in order to calculate the precession rate is the stellar gravitational quadrupole moment $J_2$.  From \citet{iorio2011classical} and \citet{ragozzine2009probing}, this is $J_2=k_2/3(q_r-q_t/2)$, where $k_2$ is the $j=2$ apsidal motion constant, $q_r=\Xi^2 R_\star^3/GM_\star$, $q_t=-3(R_\star/a)^3(M_P/M_\star)$, and $\Xi_\star=v\sin(i)/(R_{\star}\sin(\i_\star))$ is the stellar rotational angular speed. For KELT-9 $q_t$ is negligible compared to $q_r$. We take values of $k_2$ from the stellar models of \citet{claret1995stellar} with $M_\star$, $T_{\rm eff}$, and $\log g$ within the $1\sigma$ range allowed by our observations of KELT-9, finding a plausible range of $0.0022<k_2<0.0053$. Taking into account the limits on $k_2$ and 1$\sigma$ uncertainties on the measured quantities, this implies a plausible range of $5.6\times10^{-5}<J_2<2.5\times10^{-4}$.

The theoretical nodal precession rate is ${\rm d}\Omega/{\rm d}t=-J_2 3\pi/P (R_\star/a)^2 \cos(\varphi)$ \citep{barnes2013measurement,johnson2015measurement}. The likely range of $J_2$ thus corresponds to a predicted precession rate of $-0.4^{\circ} {\rm yr}^{-1}<{\rm d}\Omega/{\rm d}t<0.2^{\circ} {\rm yr}^{-1}$. Since the allowed range of $\mathbf{\varphi}$ included $\mathbf{\varphi}=90^{\circ}$ and ${\rm d}\Omega/{\rm d}t\propto\cos(\mathbf{\varphi})$, it is possible that there is no precession. In this case of $\mathbf{\varphi}=90^{\circ}$, the planetary orbit would be exactly perpendicular to the stellar equator, and so there would be no net torque on the planet from the stellar equatorial bulge to induce precession. Additionally, since the direction of precession is opposite to the direction of the orbit, the detection of precession would clarify whether or not the orbit is in fact retrograde.

\section{Discussion} \label{sec:discussion}
As the hottest discovered transiting giant exoplanet and as a gas giant in a polar orbit, KELT-9 b provides an excellent laboratory both for characterizing hot Jupiters and for understanding planet formation around high-mass stars. In the following subsections we discuss possible migration scenarios for KELT-9 b, the effect of gravity-darkening on its current-day insolation, and future work to be done on the system. 

\subsection{Possible Migration Scenarios}\label{sec:discussion.migration}
The traditional nebular hypothesis predicts that KELT-9 b should reside beyond its host star's water ice line near the system's invariable plane; however, the planet is currently in a nearly-polar 1.48-day orbit. It therefore likely migrated inward during or after its formation, and some dynamic mechanism likely caused its orbit to tilt out of alignment.

KELT-9 b likely misaligned into its polar orbit through one of three possible scenarios. One possibility is that an outside body torqued the system's protoplanetary disk out of alignment, and then KELT-9 b formed inside the misaligned plane. \citet{batygin2012primordial} and others \citep{batygin2013magnetic,lai2014star,jensen2014misaligned} demonstrated that a stellar companion can torque a disk out of the formation plane, resulting in planets already misaligned when they form. \citet{batygin2012primordial} and \citet{zanazzi2018effects} demonstrated that precession of protoplanetary disks can lead to stellar obliquity angles greater than $90^\circ$. Similarly, \citet{bate2010chaotic} and \citet{fielding2015turbulent} showed that a wide range of orbital configurations can occur when the star forms in a turbulent environment, which may have played a role in KELT-9 b's misalignment. 

Another possibility is that the host star's rotation axis torqued out of alignment. In such a scenario, KELT-9 b and any other planets in that system ostensibly remained in their formation plane, and the star instead misaligned from the system. \citet{2012ApJ...758L...6R} and \citet{rogers2013internal} demonstrate that angular momentum transport in massive stars can torque a star's envelope with respect to its rapidly rotating core, resulting in a large apparent stellar obliquity. Such a process may be detectable via asteroseismic analysis; however, following \citet{ahlers2018lasr} we do not find any evidence of stellar pulsations in KELT-9 b's \tess~photometry. 

The third general idea for explaining KELT-9 b's spin-orbit misalignment is that some mechanism misaligned the planet after formation. For example, Kozai-Lidov resonance involves bodies exchanging angular momentum by driving up inclinations and eccentricities, which could explain KELT-9 b's polar orbit \citep{fabrycky2007shrinking}.  \citet{storch2014chaotic} demonstrated that Lidov-Kozai resonance can also cause a star's rotation axis to evolve chaotically, similarly producing spin-orbit misalignment.

In addition to inclination migration, KELT-9 b likely also migrated inward during or after its formation. Several theories exist to explain the inward migration of hot Jupiters, but given KELT-9 b's polar orbit, we posit high-eccentricity migration \citep[e.g.,][]{petrovich2015hot,mustill2015destruction} as a likely cause of inward migration in this system. A dynamic event such as Lidov-Kozai resonance or scattering could have raised both KELT-9 b's eccentricity and inclination (possibly torquing KELT-9's obliquity as well), and then the planet could have recircularized via tidal dissipation, maintaining its high inclination. Ultimately, determining the cause of misalignment is beyond the scope of this work; future projects studying the dynamic behavior of this system could better constrain its migration history.

\subsection{Gravity-Darkened Seasons}\label{sec:discussion.seasons}
Throughout its orbit, KELT-9 b's received flux varies by 10\% in the ultraviolet, $1-2$\% in the visible, and less in the IR. Such a variation could have a detectable impact on KELT-9 b's overall heat transport, winds, or cloud distribution. Following Equation 4 from \citet{komacek2017atmospheric}, we estimate KELT-9 b's radiative timescale to be $\sim2$ hours at $100$ mbar and $\sim1$ day at 1 bar. Therefore, the upper atmosphere of KELT-9 b is likely changing in temperature dramatically throughout the planet's 1.48-day orbit due to its host star's gravity-darkened surface. This effect may produce strong zonal winds that could vary in intensity with the varying received stellar flux, and that could match or exceed the fast wind speeds observed on other hot Jupiters \citep[e.g.,][]{snellen2010orbital,louden2015spatially,brogi2016rotation}.

The effect of gravity-darkening on a planet's insolation can be compared to the insolation of a planet with an eccentric orbit. In both scenarios the planets receive varying amounts of flux throughout their year, which can drastically impact climate. However, the frequency of changing flux is twice per orbit for gravity-darkening versus once per orbit in eccentricity. Additionally, the effects of gravity-darkening are chromatic (with the largest flux changes typically occurring in the near ultraviolet), whereas eccentricity is achromatic. Gravity-darkening likely plays a more significant role than eccentricity for the insolation of planets such as KELT-9 b because hot Jupiter orbits are typically nearly circular.

\subsection{Future Work}\label{sec:discussion.future}
KELT-9 b is a hot Jupiter in a 1.48-day polar orbit around a bright ($V_\mathrm{mag}=7.55$) B9.5-A0 star, making it an excellent target for further study via follow-up observations. Recent studies of KELT-9 b demonstrate just how exotic these ultra-hot Jupiters can be. For example, ground-based studies by \citet{Cauley2019} and \citet{Hoeijmakers2019} have revealed the presence of metals like magnesium, iron, titanium in the extended atmosphere of KELT-9 b. Atmospheric characterization of these misaligned ultra-hot Jupiters provides constraints on the composition of their atmospheres that may in turn reveal clues to their formation history. As such,  KELT-9 b is a promising target for detailed atmospheric detections using the \emph{James Webb Space Telescope}.

KELT-9 b provides an excellent opportunity to refine previous global models tailored towards rapidly rotating stars, which will prove crucial for a large number of planets to be discovered by \tess. \citet{ahlers2020gravity} estimated that $\sim2000$ \tess~exoplanets will orbit A/F stars, many of which will be spin-orbit misaligned. Additionally, many of the host stars will be rapid rotators, placing them in regimes similar to KELT-9 b. Global model fits of these systems will yield robust, consistent estimates of the planet's bulk parameters and orbit geometries, thus better constraining the demographics of a large and interesting subset of exoplanets.

\acknowledgements
This paper includes data collected by the TESS mission, which are publicly available from the Mikulski Archive for Space Telescopes (MAST) and produced by the Science Processing Operations Center (SPOC) at NASA Ames Research Center \citep{jenkinsSPOC2016,2017ksci.rept....9J}. This research effort made use of systematic error-corrected (PDC-SAP) photometry \citep{smith2012kepler,stumpe2012kepler,stumpe2014multiscale}. Funding for the TESS mission is provided by NASA's Science Mission directorate. Resources supporting this work were provided by the NASA High-End Computing (HEC) Program through the NASA Advanced Supercomputing (NAS) Division at Ames Research Center for the production of the SPOC data products. J.P.A.’s research was supported by an appointment to the NASA Postdoctoral Program at the NASA Goddard Space Flight center, administered by Universities Space Research Association under contract with NASA. D.J.S. acknowledges funding support from the Eberly Research Fellowship from The Pennsylvania State University Eberly College of Science. The Center for Exoplanets and Habitable Worlds is supported by the Pennsylvania State University, the Eberly College of Science, and the Pennsylvania Space Grant Consortium. J.P.A. and K.D.C. acknowledge support from the GSFC Sellers Exoplanet Environments Collaboration (SEEC), which is funded in part by the NASA Planetary Science Division’s Internal Scientist Funding Model.

\bibliography{citations}
\bibliographystyle{aasjournal}

\end{document}